\documentclass[preprint,aps,showpacs,onecolumn]{revtex4} 
\usepackage{amsmath}
\usepackage{amsfonts}
\usepackage{amssymb}
\usepackage{bm}
\usepackage[dvips]{epsfig}
\usepackage[dvips]{graphicx}
\usepackage{float}
\usepackage{dcolumn}
\usepackage{ifpdf}
\usepackage{dcolumn}
\usepackage{multirow}

\newcommand{\sech}{\mathrm{sech}}
\newcommand{\be}{\begin{equation}}
\newcommand{\ee}{\end{equation}}
\newcommand{\bes}{\begin{subequations}}
\newcommand{\ees}{\end{subequations}}
\newcommand{\ben}{\begin{eqnarray}}
\newcommand{\een}{\end{eqnarray}}

\begin{document}
\title{Degenerate vacua to vacuumless model and $K\bar K$ collisions}
 \author{F. C. Simas$^{1}$, Adalto R. Gomes$^{2}$ , K. Z. Nobrega$^{3}$}
 \email{simasfc@gmail.com, argomes.ufma@gmail.com, bzuza1@yahoo.com.br}
 \noaffiliation
\affiliation{
$^1$ Centro de Ci\^encias Agr\'arias e Ambientais-CCAA, Universidade Federal do Maranh\~ao\\
(UFMA), 65500-000, Chapadinha, Maranh\~ao, Brazil\\
$^2$ Departamento de F\'\i sica, Universidade Federal do Maranh\~ao (UFMA) \\
Campus Universit\'ario do Bacanga, 65085-580, S\~ao Lu\'\i s, Maranh\~ao, Brazil\\
$^3$ Departamento de Eletro-Eletr\^onica, Instituto Federal de Educa\c c\~ao, Ci\^encia e
Tecnologia do Maranh\~ao (IFMA), Campus Monte Castelo, 65030-005, S\~ao Lu\'is, Maranh\~ao, Brazil}
\noaffiliation

\begin{abstract}
In this work we investigate a $Z_2$ symmetric model of one scalar field $\phi$ in $(1,1)$ dimension. The model is characterized by a continuous transition from a potential $V(\phi)$
with two vacua to the vacuumless case.
The model has kink and antikink solutions that minimize energy. Stability analysis are described by a Schr\"odinger-like equation with a potential that
transits from a volcano-shape with no vibrational states (in the case of vacuumless limit) to a smooth valley with one vibrational state.
We are interested on the structure of 2-bounce windows present in kink-antikink scattering processes. The standard mechanism of Campbell-Schonfeld-Wingate (CSW)
requires the presence of one vibrational state for the occurrence of 2-bounce windows. We report that the effect of increasing the separation of vacua from the potential $V(\phi)$
has the consequence of trading some of the first 2-bounce windows predicted by the CSW mechanism by false 2-bounce windows. Another consequence is the appearance of false 2-bounce windows of zero-order.

\end{abstract}


\maketitle


\section{ Introduction }
Solitary waves are solutions from nonlinear physics characterized by the very special property
of localized density energy that can freely propagate without loosing form.
The realization of solitary waves in nature is now amply recognizable, with interesting effects in solid state and atomic physics \cite{dauxois}. In cosmology solitary waves are also studied in the context of bubble collisions in the primordial universe, where in some limit the effect of gravitational interaction between the bubbles can be negligible and the process of collisions can be described by an effective $(1,1)$ dimensional model \cite{bubbles}. The simplest solitary wave is described by the $(1,1)$ dimensional kink. Usually kink models are constructed starting from a Lagrangian density
\begin{eqnarray}
{\cal{L}} = \frac12 \partial_{\mu} \phi \partial^{\mu} \phi - V(\phi),
\end{eqnarray}
where the $\phi$ is a real scalar field and a potential $V(\phi)$ with minima at non-zero values of $\phi$ that suffers a spontaneous symmetry breaking. The equation of motion for the field $\phi(x,t)$ is given by
\begin{eqnarray}
\label{eom}
\phi_{tt}-\phi_{xx}+V_{\phi}=0,
\end{eqnarray}
where $V_{\phi}=dV/d\phi$. If we can write the potential in terms of the superpotential $V(\phi)=\frac12W_{\phi}^2$, we note that the energy will obeys the first-order differential equation
\begin{eqnarray}
\frac{d\phi}{dx}=W_{\phi}.
\end{eqnarray}
The defects formed with this prescription minimize energy and are known as BPS defects \cite{bps1, bps2}.

Stability analysis considers small
fluctuations
\be
\Phi(x,t)=\phi(x)+\eta(x)e^{i\omega t},
\ee
resulting in a Sch\"odinger-like equation
\be
-\frac{d^2\eta}{dx^2}+V_{sch}(x)\eta=\omega^2\eta
\ee
with
\be
V_{sch}(x) = V_{\phi\phi}(\phi(x)).
\ee

As one knows\cite{bazeia}, this equation is factorizable, since it can be written as
\be
S_\pm^\dagger S_\pm \eta =\omega^2\eta
\ee
with
\be
S_\pm=-\frac d {dx} \pm W_{\phi\phi},
\ee
which forbids the existence of tachionic modes. Zero-mode exist, corresponding to a translational degree of freedom of the kink/antinink,  and is given by
\be
\eta_0=CW_\phi,
\ee
with $C$ a normalization constant.

Nonintegrable models like those considered here have the scattering process with a very rich character. The archetype model is the $\phi^4$, where
the main aspects where studied in deep (see for instance refs. \cite{w1,aom,gh}). There one knows that large initial velocities $v>v_{crit}$  lead to a simple scattering process where the $K \bar K$ pair encounter and after a single contact recede from themselves. This is called a 1-bounce process. Small initial velocities lead to the formation of a bound $K \bar K$ state called bion that radiate continuously until be completely annihilated. In the bion region and near to the frontier region in velocities $v\sim v_{crit}$ it can occur the formation of 2-bounce windows that accumulates toward $v= v_{crit}$ with lower and lower thickness. Between a border of a 2-bounce windows and the bion region another system of 3-bounce windows can be found. This substructure is verified for even high levels of bounce windows in a fractal structure.

Stability analysis of the kink lead to a Schr\"odinger-like equation where the presence of zero-mode in related to the translational invariance of the kink (invariance under Lorentz boosts).
Usually one interprets the presence of non-null bound states as vibrational states. In the standard Campbell-Schonfeld-Wingate (CSW) mechanism, a resonant exchange of
 energy between the translational and vibrational modes is responsible for the structure of 2-bounce windows\cite{csw}. As far as we know there are two exceptions
to this scenario: i) despite the absence of vibrational mode in the perturbation of a kink, considering the effect of collective $\bar K K$ structure, it was
explained the occurrence of 2-bounce windows in the $\phi^6$ model\cite{phi6}; ii) the presence of more that one vibrational state can in some circumstances
destroy the 2-bounce structure\cite{sgno}.

In this work we are interested in study the effect of the separation of the vacua of the potential $V(\phi)$ in the process of $K\bar K$ collision, focusing mainly on the appearance and structure of 2-bounce windows.
 In the Sect. II we will investigate a model with unusual scattering properties that can also contribute for the
understanding of the mechanism of formation of 2-bounce windows. In Sect. III we present the numerical analysis of the $K\bar K$ scattering process.
 Our main conclusions are reported in Sect. IV.

\section{The model}

An exception of the known mechanism for constructing kinks are the so called vacuumless defects which are constructed in models with a potential that has a local  maximum but no minima.
One example is the model proposed by Cho and Vilenkin\cite{ChoV}:
\be
V(\phi)=\sech^2(\phi),
\ee
Potentials of this type appear also in non-perturbative effects in supersymmetric gauge
theories\cite{susy}.
 The equation of motion for the scalar field has the solution
\be
\phi(x)=\sinh^{-1}(\sqrt2 x)
\ee
which has finite energy.

Static solutions of this model where further studied both in their gravitational aspects in $(3,1)$ dimensions \cite{ChoV2} and concerning to their topological structure and trapping of fields in $(1,1)$ dimensions \cite{DB1999}. More recently, Dutra and Faria Jr\cite{SDAF} considered an extension described by the potential
\begin{eqnarray}
\label{V}
V(\phi) = \frac12 \bigg(A\cosh(\phi) - \sech(\phi) \bigg)^2.
\end{eqnarray}
\begin{figure}
\includegraphics[{angle=0,width=6cm,height=6cm}]{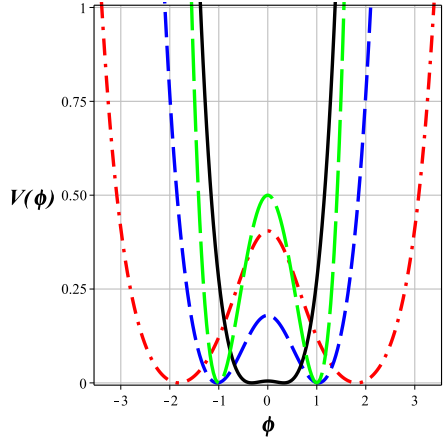}
\caption{(a) Potential $V(\phi)$. The figures are for fixed $A=0.1$ (red dash-dotted), $A=0.4$ (blue dash), $A=0.9$ (black line). Here it is also representing the potential
 for the $\phi^4$ model (long-dashed green).}
\label{fig_1}
\end{figure}

Figure 1 shows plots of $V(\phi)$ for several values of $0<A<1$, where there is the presence of two symmetric minima (due to the $Z_2$ symmetry) and a local maximum at $\phi=0$.
When $A$ is reduced, the vacua are located at larger values of $|\phi|$ and the local maximum grows. The
vacuumless potential from Cho and Vilenkin is recovered for $A\to0$. Only for comparison we plot the potential for the $\phi^4$ model, showing that for the same vacua ($\phi=\pm1$), the local maxima of
the $\phi^4$ potential is higher.

\begin{figure}
\includegraphics[{angle=0,width=6cm,height=6cm}]{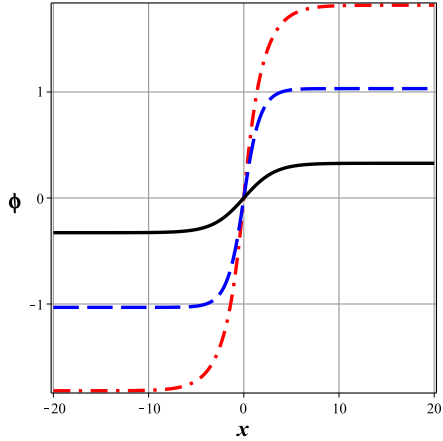}
\caption{Field configurations $\phi(x)$. The figures are for fixed $A=0.1$ (red dash-dotted), $A=0.4$ (blue dash), $A=0.9$ (black line).} \label{fig_1}
\end{figure}
Static solutions for the scalar field are
\begin{eqnarray}
\phi(x)=\pm \sinh^{-1}{\bigg(\sqrt{\frac{1-A}{A}}\tanh \big(\sqrt{A(1-A)} x \big)\bigg)},
\end{eqnarray}
where plus and minus signs are for kink ($K$) and antikink ($\bar K$), respectively. The vacua of the model are described by \cite{SDAF}
\begin{eqnarray}
\phi(x\to \pm\infty)=\pm \cosh^{-1}\bigg(\sqrt{\frac1A} \bigg).
\end{eqnarray}
Figure \ref{fig_1} depicts some plots of $\phi(x)$ for several values of $A$. Note from the figure that the increasing in the separation of the vacua turns the kinks ticker.
In the limit of a vacuumless model $A=0$ the maximum of $\phi(x)$ goes to infinity. Despite this the energy density is a localized function of $x$ even in the vacuumless case (for more
details see ref. \cite{SDAF}).

\begin{figure}
\includegraphics[{angle=0,width=6cm,height=6cm}]{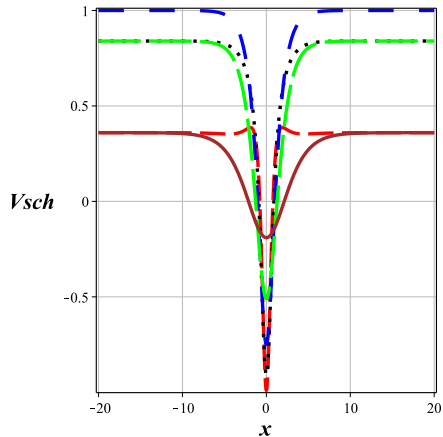}
\caption{Schr\"odinger-like potential $V_{sch(x)}$. The figures are for fixed $A=0.1$ (red dash), $A=0.3$ (black dotted), $A=0.5$ (blue spacedash), $A=0.7$ (green long-dash) and $A=0.9$ (brown line).}
\label{fig_pot}
\end{figure}

Figure \ref{fig_pot} depicts some plots of the Schr\"odinger-like potential $V_{sch}$ of perturbations for several values of $A$.
 When $A\to 0$, we have a the perturbation potential with a volcano-shape with a deep minimum. The increasing
of $A$ up to $A=0.5$ reduces the depth of the minimum and increases the asymptotic maximum of the potential. For $0.5<A<1$ the asymptotic maxima is reduced whereas the depth of the minimum continues
to decrease. Then in this region we have a continuous reduction of the difference between minimum and maximum of the potential, with a broader shape.

\begin{table}[tbp]\scriptsize
\begin{tabular}{|c|c|c|c|c|c|c|c|c|c|c|c|c|c|}
\hline
$A$                      & 0.1       & 0.2    & 0.3     & 0.4    & 0.5    & 0.6    & 0.7     & 0.8    & 0.9  & 0.98          \\ \hline
$\omega^2$ for vibrational state             & -         & -      & 0.8256  & 0.9006 & 0.8946 & 0.8223 & 0.6920  & 0.5093 & 0.2777  & 0.0609   \\ \hline
order $m$ of first 2-bounce window     & -         & -      & -       & 3  & 1     & 1     & 1     & 1    & 1   & 1     \\ \hline
\end{tabular}
\caption{Structure of vibrational states with $A$.}
\end{table}

With this potential the occurrence of bound states was investigated numerically. For all values of $0<A<1$ there is always a zero-mode.
The structure of vibrational states is summarized in Table I. There one can see that for small values of $A$ only a zero-mode appears. In particular, case $A=0$, the true vacuumless case,
 which has a volcano-shape potential $V_{sch}$, possibly has  resonances. However, since the existence of resonances is not relevant for the formation of 2-bounces we will not elaborate on this.
For $0.3 \lesssim A \lesssim 1$ we have the presence of one vibrational mode. Note also that the energy of the vibrational mode grows with $A$ until a maximum for $A\sim 0.4$, reducing monothonically for
larger values of $A$ and going to zero for $A\to 1$.

\section { Numerical Results }

In this section we present our main results of the $K\bar K$ scattering process. We are interested in the effect of the separation of vacua of $V(\phi)$ given by
Eq. (\ref{V}) for intermediate initial velocities, where it is verified the presence of bounce windows. In particular we focus in the structure of 2-bounce windows
 and their relation with the structure of vibrational states.

 We considered a symmetric $K\bar K$ collision. Initial conditions are given by sufficiently distant defects described by boosted free solutions:
\begin{eqnarray}
\phi(x,0) &=& \phi_{K}(x+x_0,0)-\phi_{\bar{K}}(x-x_0,0)-\cosh^{-1}\bigg(\sqrt{\frac1A} \bigg), \\
\dot\phi(x,0) &=& \phi_{K}(x+x_0,0)-\dot\phi_{\bar{K}}(x-x_0,0). \label{initial_cond} \end{eqnarray}
Here $\phi_K$ means free kink solution. We fixed $x_0=15$ as the initial kink position, i.e., the kink solution centered at $-x_0$ and the antikink at $x_0$.
In this work, for solving the equation of motion given by Eq. (\ref{eom}) we used a pseudospectral method on a grid with $2048$ nodes with periodic boundary conditions for $\phi$ and $\dot\phi$ and we set the grid boundary at $x_{max}=200$.

\begin{figure}
\includegraphics[{angle=0,width=6cm}]{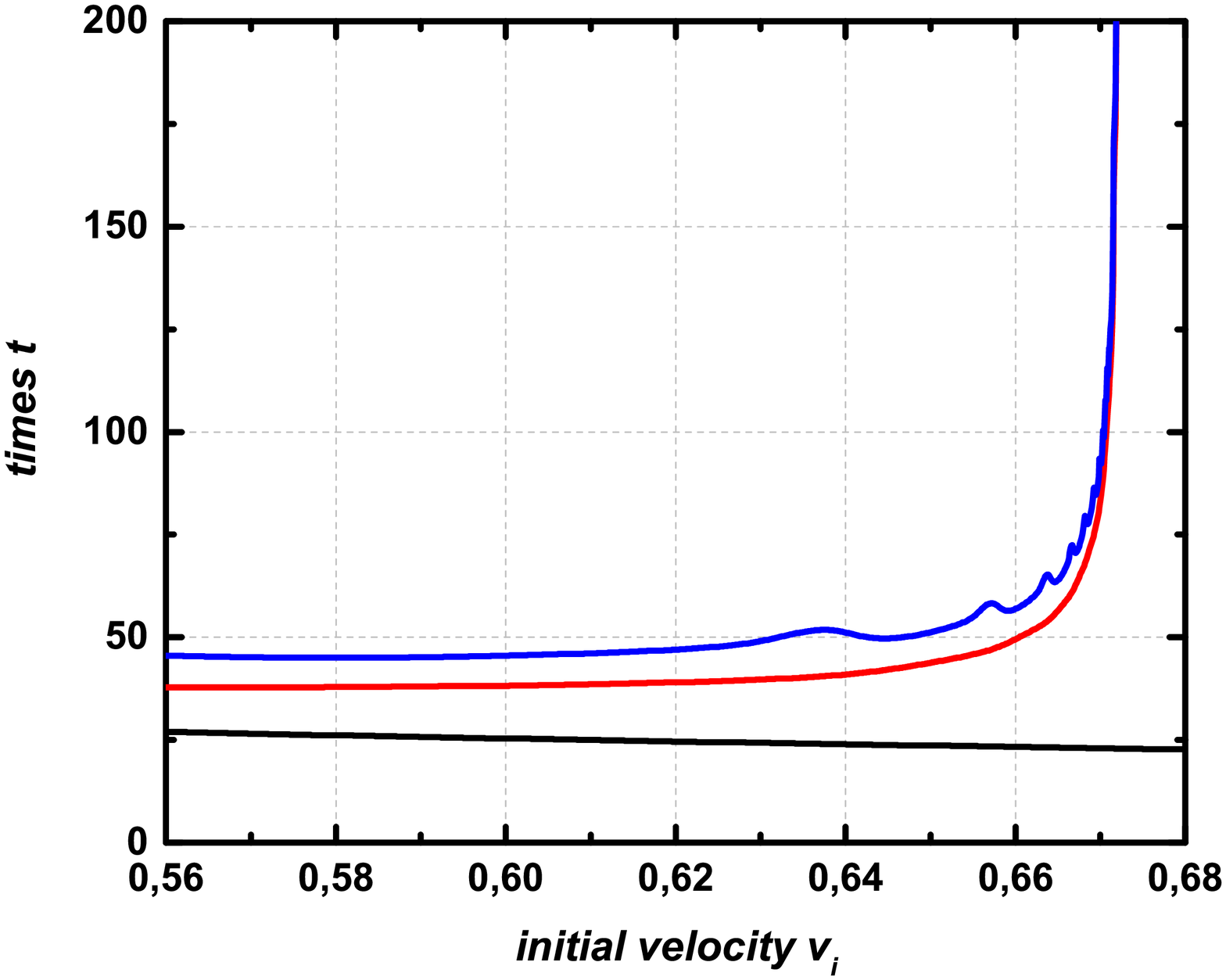}
\includegraphics[{angle=0,width=6cm}]{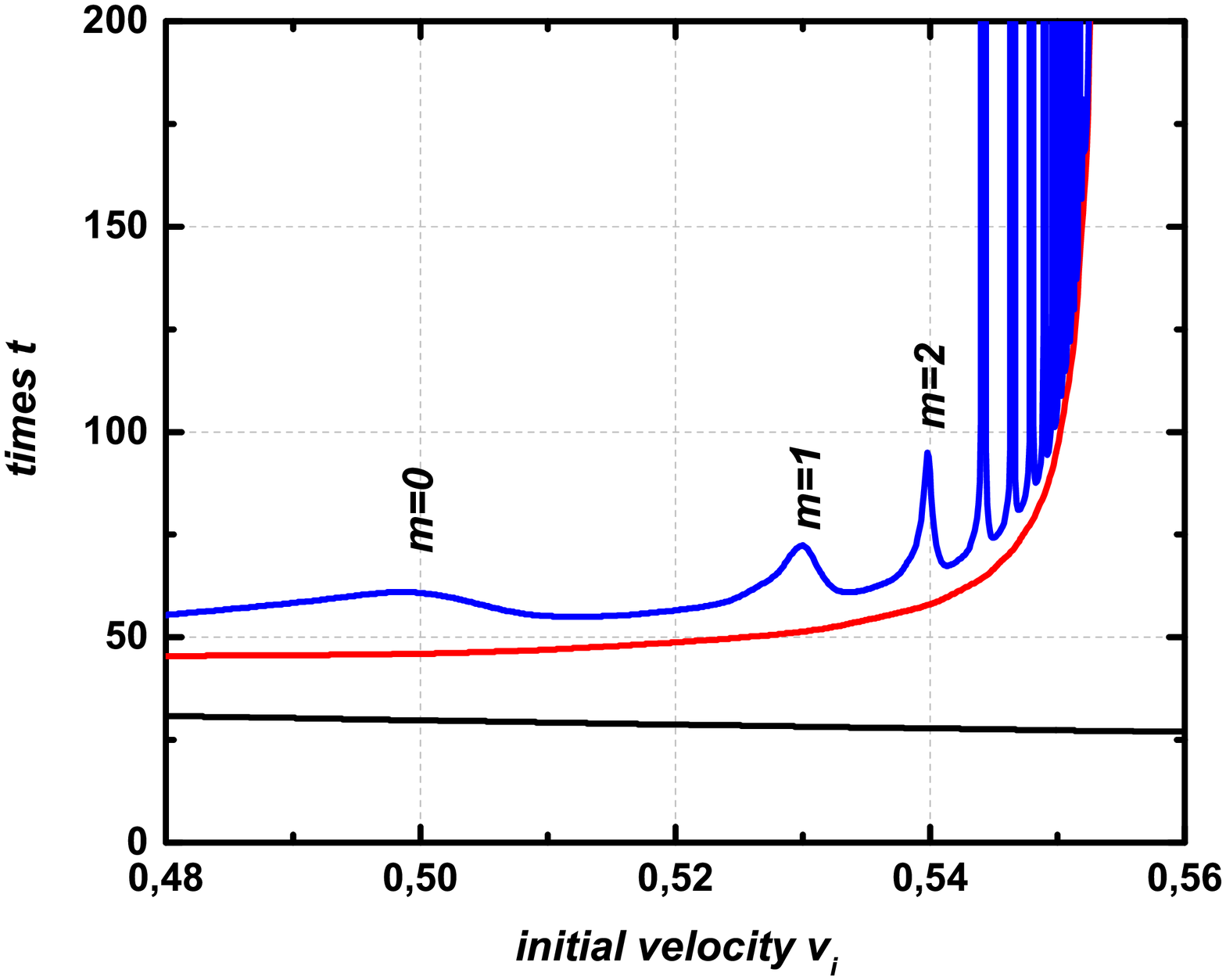}
\includegraphics[{angle=0,width=6cm}]{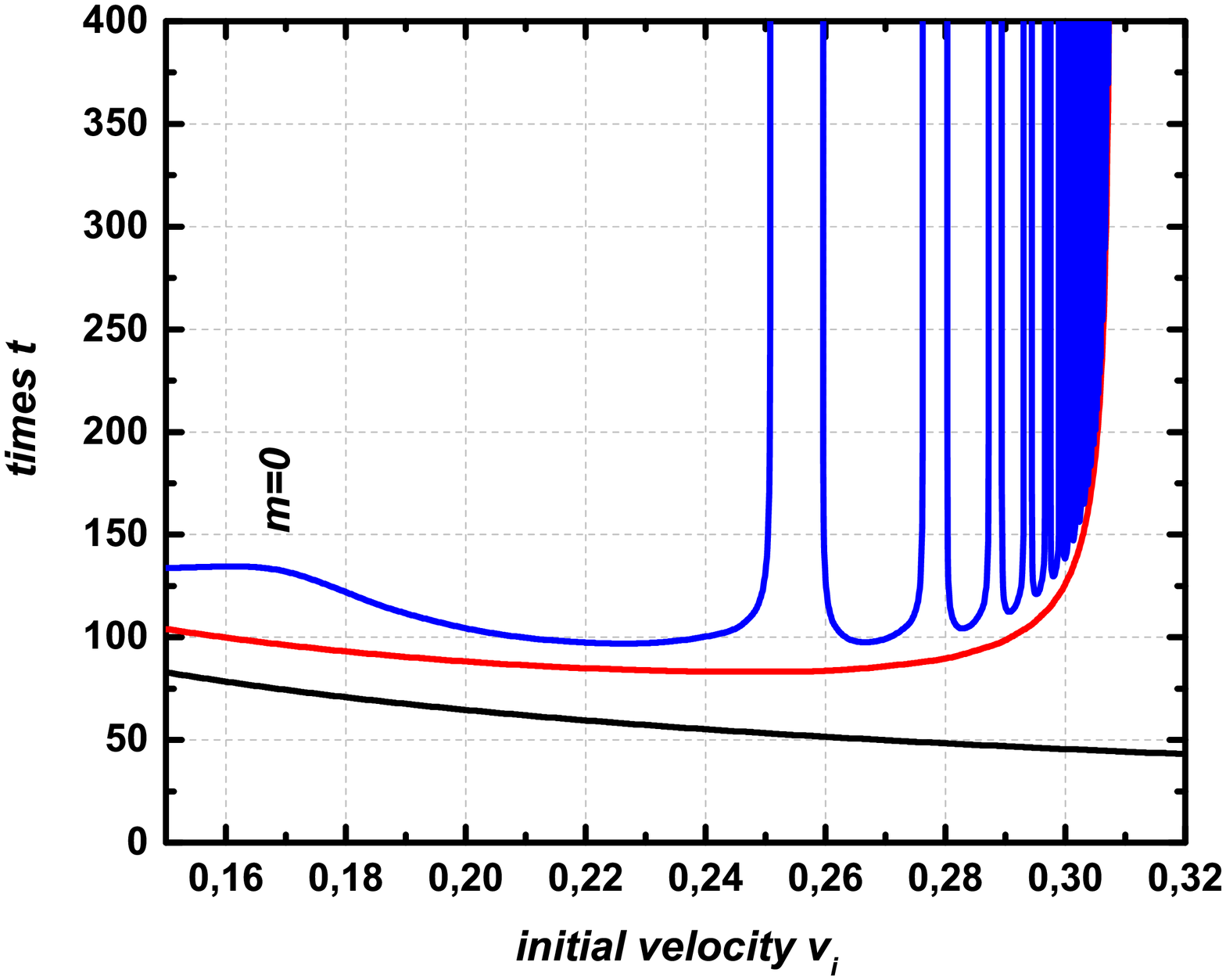}
\includegraphics[{angle=0,width=6cm}]{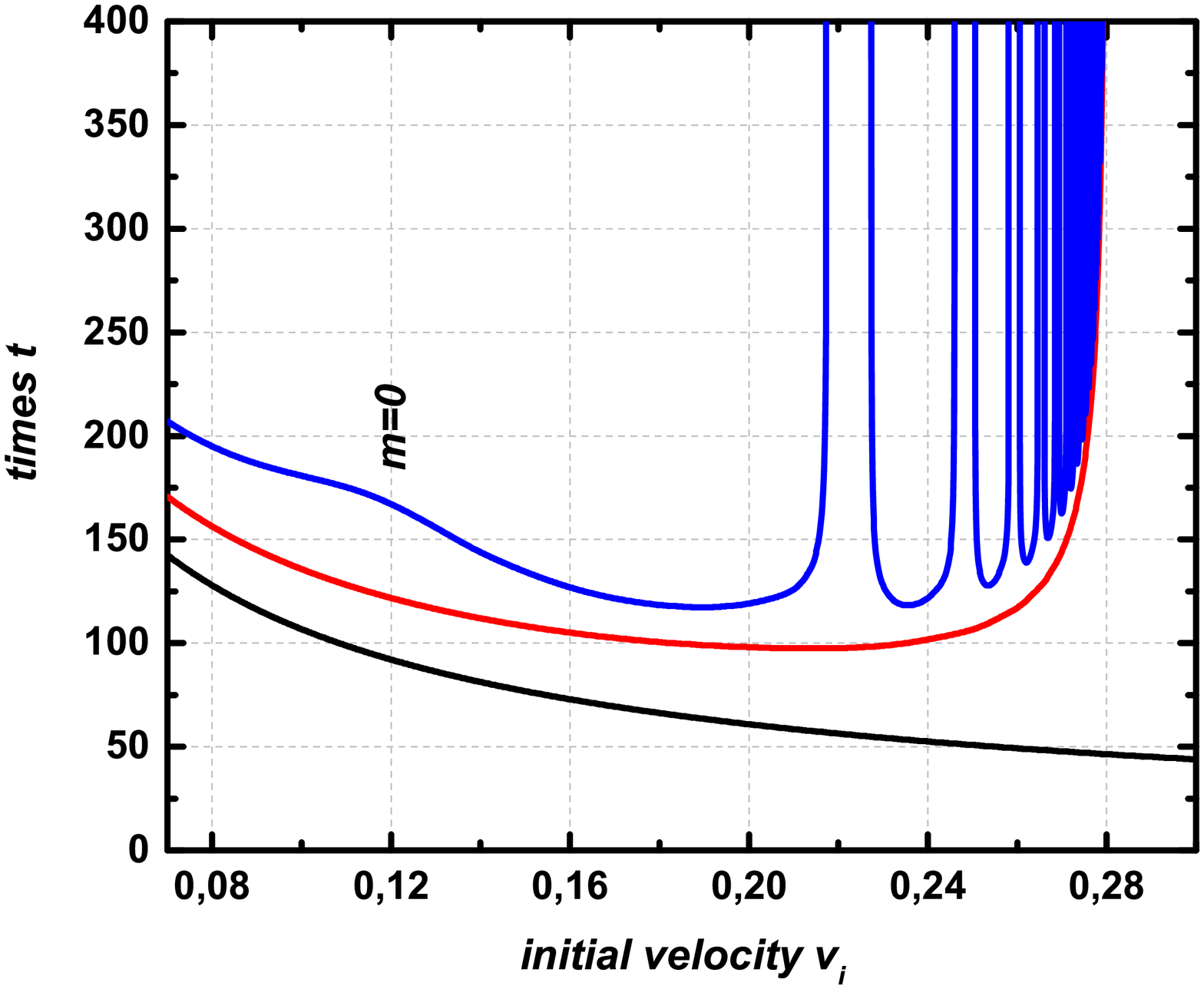}
\caption{Times to the first (black), second (red) and third (blue) kink-antikink collisions as a function of initial velocity $v_i$ for  a) $0.3$, b) $0.4$, c) $0.8$, d) $0.9$.}
\label{fig_2}
\end{figure}

Small values of  $A$  ($A\lesssim 0.2$) are cases without vibrational mode, where there is no 2-bounce windows at all,
in accord with what expected from the standard CSW mechanism\cite{csw}. Note that for this case the critical velocity $v_{crit}$ is close to 1, showing that the scattering process
 leads almost always to bion states with just a small interval in initial velocities leading to collisions with 1-bounce. We also noted that $v_{crit}$ grows with the reduction
of $A$ showing that in the limit of a vacuumless model only bion states are produced.

The structure of 2-bounce windows for larger values of $A$ ($0.3\lesssim A <1$) can be revealed in Figs. \ref{fig_2}a-\ref{fig_2}d where we plot the behavior of the times of first, second and third kink-antikink collisions as a function of initial velocity for fixed values of $A$. As showed in Table I, these cases correspond to examples with one vibrational mode. For $A=0.9$ (Fig. \ref{fig_2}d), the presence of a vibrational mode induce the appearance of 2-bounce windows, recognized in the figure by regions where there is divergence
in the times corresponding to a third collision.  We note also the small critical velocity $v_{crit}\sim 0.28$ (when compared to $v_{crit}\sim 0.97$ for $A=0.1$), indicating that
 potentials with degenerated vacua with smaller $|\phi|$ favor the enlargement of the 1-bounce region. We note also the presence of a first smooth peak unexpected from CSW mechanism. This is called zero-order false 2-bounce windows, and where first observed in the context of $\phi^6$ model\cite{phi6}.

Now if we slowly reduce the value of $A$ back to zero,  the structure of zero-order false 2-bounce windows
is enlarged, reducing their height until finally disappear for $A\lesssim 0.2$. Also for $A\sim 0.4$ there is the formation of two more false 2-bounce windows beside the zeroth-order.
This is possibly connected to the presence of a maximum in the energy of the vibrational state, as seen in Table I. This behavior of false windows grows until a complete disappearance of 2-bounce windows
 in Fig. \ref{fig_2}a for $A=0.3$, substituted by bion and 1-bounce scattering, despite the presence of a vibrational state. There, instead of the expected windows, we have a set
of oscillations accumulating in $v=v_{crit}$. We note also that the larger is $A$, the larger is the thickness of the 2-bounce windows (compare Fig \ref{fig_2}b for $A=0.4$ with Fig. \ref{fig_2}d for $A=0.9$).

\begin{figure}
\includegraphics[{angle=0,width=5cm}]{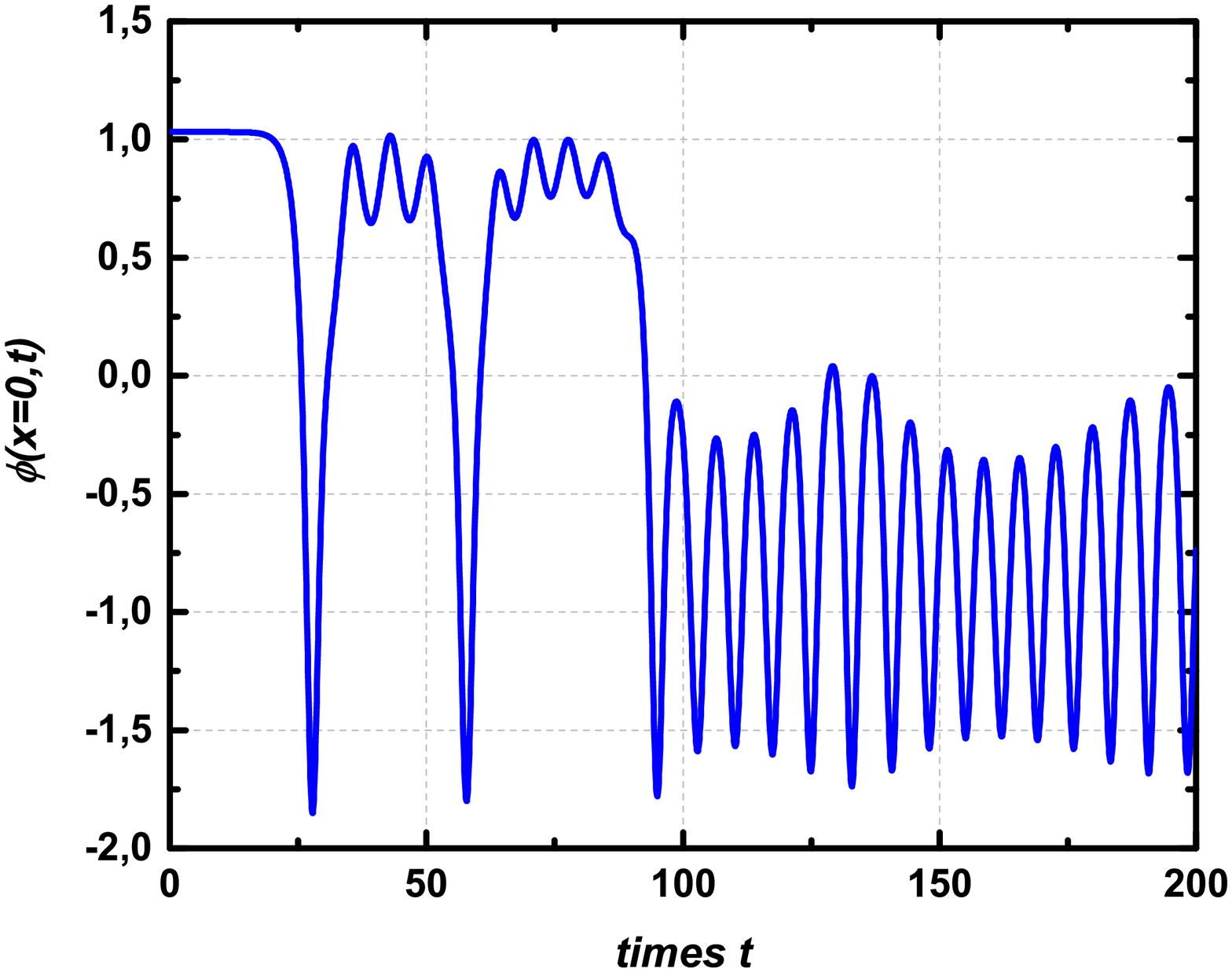}
\includegraphics[{angle=0,width=5cm}]{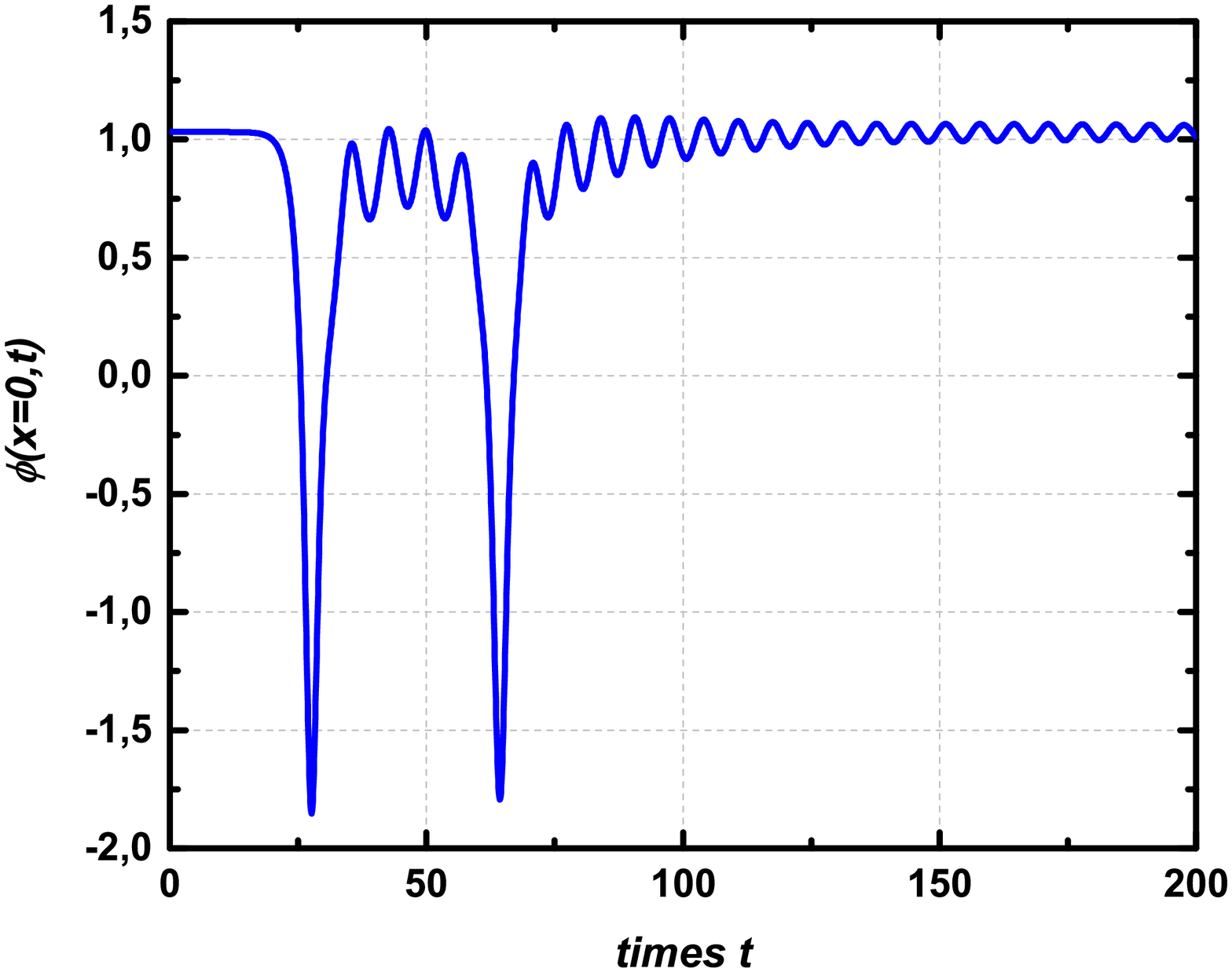}
\includegraphics[{angle=0,width=5cm}]{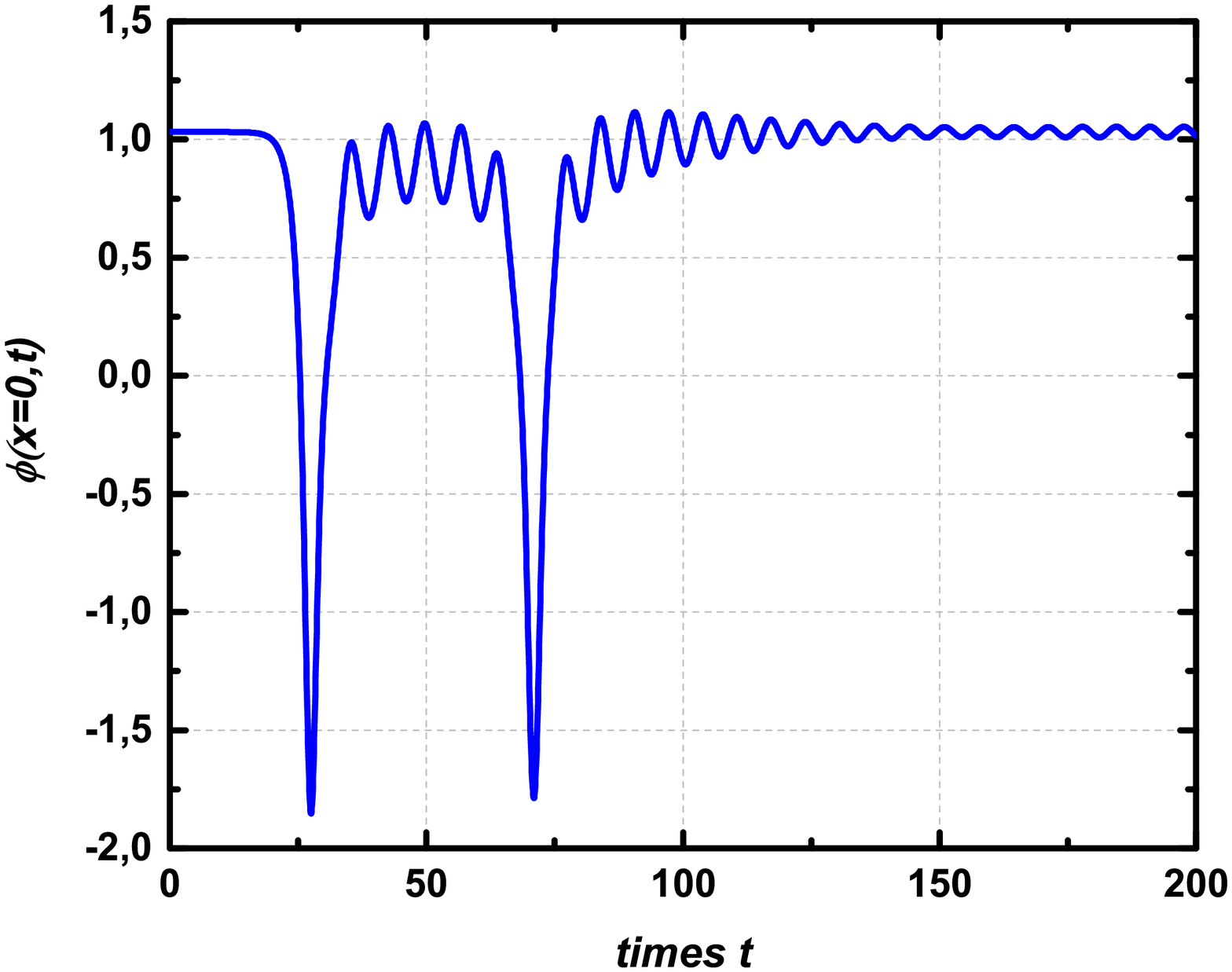}
\caption{Scalar field at the center of mass $\phi(0,t)$ versus time for for fixed $A=0.4$ and (a) $v_i=0.5398$, (b) $v_i=0.5441$ and (c) $v_i=0.5464$. }
\label{fig_3}
\end{figure}

The peaks corresponding to false windows and true 2-bounce windows can be analyzed counting the number $M$ of cycle oscillations between each collision. Each window can be labeled
by an integer $m=M-2$\cite{aom}. In Figs. \ref{fig_3}a-\ref{fig_3}c we describe the behavior of $\phi(0,t)$ for case $A=0.4$, corresponding to the diagram $t\times v_i$ of  Fig. \ref{fig_2}b.
 Figs. \ref{fig_3}b and \ref{fig_3}c show that $\phi(0,t)$ oscillates $M=5$ and $M=6$ times, respectively, corresponding to the expected third ($m=3$) and fourth ($m=4$)
2-bounce windows. The first two 2-bounce windows are missed. One example from the second ($m=2$) missed 2-bounce windows is seen in Fig. \ref{fig_3}a,
 corresponding to the third peak of Fig. \ref{fig_2}b. Indeed including the first two bounces of Fig. \ref{fig_3}a the function $\phi(0,t)$ oscillates $M=4$ times, so the
 order of the peak is $m=2$.  The structure of false 2-bounce windows is evident since after the second bounce $\phi(0,t)$ oscillates for a while around the vacuum ($\phi=1$,
 according to Fig. \ref{fig_1}), decaying further to the second vacuum, with the field oscillating in a bion pattern around $\phi\sim-1$.

\section { Conclusions  }

In this work we have studied a hybrid degenerate vacua to vacuumless model depending on a parameter $A$. We found that the separation of vacua have sensible effects in the structure of 2-bounce windows.
It was shown that the potential of perturbations transits from a smooth valley (with one vibrational state) to a volcano shape (with no vibrational state). The interesting region is that of an intermediate value of $A=0.4$, where the energy of the vibrational state is maximum. In this region the structure of 2-bounce windows starts to be traded by false windows. The false 2-bounce windows are characterized by a $K\bar K$ scattering process where the 2-bounces occur, but instead of being separated, the pair is turned either to an oscillon or to a bion state, finally resulting in total anihilation. In particular, the oscillon state seems to have more general relation to the nonlinearity of the model, having been observed for some velocities even far from the 2-bounce windows (true of false).

This process continues when one reduces the parameter to the region where the vacua are more departed. In particular case $A=0.3$ is characterized by the total suppression of 2-bounce windows, substituted by a sequence of false 2-bounce windows, despite having one vibrational state.  Here the CSW mechanism works in the sence of predicting the existence and characteristics of the scattering processes from each 2-bounce windows. However, despite the initial scattering mechanism occurs accordingly, in the end it is frustrated, the pair being not allowed to separate anymore. In some sense the model introduces an asymmetry in the influence of $A$ on the behavior of the scattering process: region of $A<A_c=0.4$ (where $A_c$ marks the maximum of energy of vibrational state) favors the frustration of the scattering mechanism followed by a changing of vacuum of the solutions. On the other hand, region $A>A_c$ there is no such process. This is a clear indication of the influence of a vacuumless character on the frustration of 2-bounce windows.


\section{Acknowledgements}
The authors thank FAPEMA and CNPq for financial support.


\end{document}